\documentclass[letterpaper,conference]{IEEEtran}
\usepackage{hyperref}

\usepackage{xcolor}
\usepackage{listings}
\usepackage{inconsolata}
\usepackage{relsize}
\usepackage{amsmath,amssymb,array}
\usepackage{esz}
\usepackage{mathpartir}
\usepackage{graphicx}
\usepackage{xspace}
\usepackage{xurl}
\usepackage{caption}
\usepackage{bbding}
\usepackage{tabularx}
\usepackage{enumitem}

\hypersetup{colorlinks=true,linkcolor=blue,citecolor=blue,urlcolor=blue}

\newcommand{\Section}[1]{\section{#1}}
\newcommand{\SubSection}[1]{\subsection{#1}}
\newcommand{\Paragraph}[1]{\vspace{0.5ex}\noindent\textbf{#1.}\xspace}

\newlist{Itemize}{itemize}{1}
\setlist[Itemize]{label=--, leftmargin=1em, labelindent=0pt, labelsep=0.5em, itemsep=0.5ex, topsep=0.5ex}

\newenvironment{WideTable}{
    \begin{table*}[!t]
} {
    \end{table*}
}

\newcommand{\MVP}[0]{\textsf{MVP}\xspace}

\newcommand{\WP}[0]{\textsf{WP}\xspace}
\newcommand{\solidity}[0]{\textsf{Solidity}\xspace}

\newcommand{\fstar}[0]{\textsf{F*}\xspace}
\newcommand{\verus}[0]{\textsf{Verus}\xspace}
\newcommand{\certora}[0]{\textsf{Certora}\xspace}

\newenvironment{ivl}{\zedindent=2ex\begin{zed}}{\end{zed}}

\newcommand{\requiresof}[2]{\Zpreop{requires\_of}\langle#1\rangle(#2)}
\newcommand{\abortsof}[2]{\Zpreop{aborts\_of}\langle#1\rangle(#2)}
\newcommand{\ensuresof}[2]{\Zpreop{ensures\_of}\langle#1\rangle(#2)}
\newcommand{\resultof}[2]{\Zpreop{result\_of}\langle#1\rangle(#2)}

\newcommand{\Clo}{\mathcal{C}}
\newcommand{\Par}{\mathcal{P}}
\newcommand{\Fld}{\mathcal{F}}

\newcommand{\Addr}{\Zpreop{address}}

\newcommand{\Mem}{\mathcal{M}}
\newcommand{\Mod}{\Delta}
\newcommand{\Type}{\mathcal{T}}
\newcommand{\Expr}{\mathcal{E}}

\newcommand{\PROC}{\Zkeyword{proc}}
\newcommand{\FUN}{\Zkeyword{fun}}
\newcommand{\AXIOM}{\Zkeyword{axiom}}

\newcommand{\RETURNS}{\Zinrel{returns}}
\newcommand{\HAVOC}{\Zkeyword{havoc}}
\newcommand{\ASSUME}{\Zkeyword{assume}}
\newcommand{\ASSERT}{\Zkeyword{assert}}
\newcommand{\DATATYPE}{\Zkeyword{datatype}}
\newcommand{\IS}{\Zinrel{is}}
\newcommand{\TRUE}{\Zkeyword{true}}

\newcommand{\BOOL}{\Zpreop{bool}}

\newcommand{\CALL}{\Zpreop{call}}

\newcommand{\sat}{\mathrel{{\scriptstyle\vert}\mkern-1mu{\scriptstyle\sim}}}
\newcommand{\slabeled}[2]{#1 \sat #2}

\newcommand{\sq}[1]{\overline{#1}}

\lstdefinestyle{MoveStyle}{
    basicstyle=\ttfamily,
    keywordstyle=\color{black}, 
    commentstyle=\color{gray}\normalfont\itshape,
    escapechar=@, 
    breaklines=true,
    literate={|~}{{$\sat$\,}}1
        {<=}{{$\leq$}}2
        {>=}{{$\geq$}}2
        {==}{{$=$}}2
        {!=}{{$\neq$}}2
        {==>}{{$\Longrightarrow$}}3
        {forall}{{$\forall$}}1
        {exists}{{$\exists$}}1,
}

\lstdefinelanguage{Move}{
    morekeywords={
        abort,
        aborts_if,
        aborts_of,
        acquires,
        address,
        as,
        assert,
        assume,
        borrow_global,
        borrow_global_mut,
        break,
        const,
        continue,
        copy,
        copyable,
        define,
        drop,
        else,
        ensures,
        ensures_of,
        exists,
        false,
        forall,
        friend,
        fun,
        global,
        has,
        havoc,
        if,
        in,
        include,
        invariant,
        key,
        let,
        loop,
        match,
        modifies,
        modifies_of,
        module,
        move,
        move_from,
        move_to,
        mut,
        native,
        num,
        old,
        onabort,
        pragma,
        proof,
        public,
        requires,
        requires_of,
        resource,
        result_of,
        return,
        schema,
        script,
        signer,
        spec,
        split,
        store,
        struct,
        true,
        u8,
        u64,
        u128,
        u256,
        update,
        use,
        with,
        where,
        while},
    sensitive=true,
    morecomment=[l]{//},
    morecomment=[s]{/*}{*/},
}

\lstMakeShortInline[language=Move,style=MoveStyle]|
\lstMakeShortInline[language=Move,style=MoveStyle]!

\lstnewenvironment{Move}{
    \lstset{
        language=Move,
        style=MoveStyle,
        basicstyle=\relsize{-1}\ttfamily,
        keywordstyle=\color{blue},
    }
}{
}

\lstnewenvironment{MoveBox}{
    \lstset{
        language=Move,
        style=MoveStyle,
        basicstyle=\relsize{-1}\ttfamily,
        keywordstyle=\color{blue},
    }
}{
}

\lstnewenvironment{MoveBoxNumbered}{
    \lstset{
        language=Move,
        style=MoveStyle,
        basicstyle=\relsize{-1}\ttfamily,
        keywordstyle=\color{blue},
        numbers=left,
        numberstyle=\scriptsize\color{gray},
    }
}{
}

\lstnewenvironment{MoveDiag}{
    \lstset{
        style=MoveStyle,
        basicstyle=\relsize{-2}\ttfamily,
    }
}{
}

\begin{document}

\title{Formal Verification of Imperative \\First-Class Functions in Move}

\author{%
	\IEEEauthorblockN{Wolfgang Grieskamp\IEEEauthorrefmark{1} and Teng Zhang
	 and Vineeth Kashyap and Jake Silverman}
	\IEEEauthorblockA{Aptos Labs, Palo Alto, USA\\
		\texttt{\{wg,teng,vineeth,jake.silverman\}@aptoslabs.com}\\
		\IEEEauthorrefmark{1}Corresponding author}
}

\maketitle

\bstctlcite{BSTcontrol}

\begin{abstract}
	The Move Prover (\MVP) is a formal verifier for smart contracts written in the Move programming language.
	Recently, Move on Aptos was extended with \emph{higher-order functions}: imperative functions as first-class values that can be passed around, stored in data structs, and kept in persistent storage, enabling \emph{dynamic dispatch}.
	This paper describes the representation of function values in the Move specification language and their implementation in \MVP.
	We introduce \emph{behavioral predicates} which characterize Move functions (aborts and pre-/postconditions) by single-state or two-state predicates.
	We also introduce \emph{state labels} for naming intermediate memory states in which expressions are evaluated and which allow to compose behavioral predicates to describe sequences of state transitions.
	On SMT level, function values are encoded by discriminating over the possible function values reaching a call site: when the concrete function is known, its effect is accounted for directly; when it is unknown (for example, a function parameter, or a closure loaded from storage), its behavioral predicates describe the effect.
    Our approach goes beyond, for example, Dafny, by supporting imperative first-class functions which can modify state via Rust-style references and global variables, and leads to more efficient SMT encodings than separation logic because of the static separation of memory enabled by Move.
	We further extend \MVP's \emph{specification inference} tool to work with function values: given arbitrary higher-order Move code, weakest-precondition analysis semi-automatically derives behavioral-predicate-based specifications, reducing the annotation burden and providing a validation pipeline for the new specification constructs.
\end{abstract}

\Section{Introduction}
\label{sec:intro}

The Move Prover (\MVP)~\cite{DillGPQXZ22} is a formal verification tool for Move smart contracts~\cite{blackshear2020resources,MoveOnAptosBook}, designed for routine use during development with response times comparable to type checkers and linters.
\MVP is deployed at Aptos, a public layer-1 blockchain whose smart contracts are written in Move, where it verifies core protocol logic --- staking, metering, code deployment, and supporting data structures~\cite{ParkZGXGCLC24, AptosFrameworkBook}; it was originally developed for the Libra/Diem blockchain at Meta.

Recently, Move was extended with \emph{higher-order functions}~\cite{grieskamp2025movehigherorder}: functions are now first-class values that can be passed, returned, stored in struct fields, and persisted in global storage.
Combined with Move's resource model, persisted function values give a principled account of \emph{dynamic dispatch}.
Typical Aptos use cases include callback-based asset flows, extensible asset managers, dispatchable permissions and metering, and generic data-structure operations parameterized by a caller-provided function (e.g.\ fold, map, filter, custom comparators).

Higher-order functions are a known challenge for verification: they break the modular, static-dispatch specification style on which \MVP was built.
When a call is dispatched dynamically, its pre- and post-conditions cannot in general be looked up at the call site; the caller's specification must instead refer \emph{abstractly} to the behavior of its function-value arguments, and the verifier must reconcile this with whichever concrete function reaches the call site.
For a production-oriented prover, this reconciliation must be fast, complete in the relevant cases, and free of the trigger instability (`butterfly' effect) that SMT encodings easily introduce~\cite{BUTTERFLY}.

\Paragraph{Contributions}
This paper describes the extension of \MVP to support higher-order Move, and the underlying extension of the Move specification language:

\begin{Itemize}
    \item We extend Move specifications with \emph{behavioral predicates} $\requiresof{f}{\bar{x}}$, $\abortsof{f}{\bar{x}}$, and $\ensuresof{f}{\bar{x}, \bar{y}}$, which characterize the pre-/abort-/post-behavior of a function value $f$ on arguments $\bar{x}$ and results $\bar{y}$, making function specifications first-class. They implicitly track read and read/write state dependencies.

    \item We introduce \emph{state labels}, which name intermediate memory states along a function's execution and connect them via two-state predicates, as in $\slabeled{S1..S2}{\phi}$, enabling reasoning about sequential evolution of state.

    \item We describe how \MVP implements invocation of function values on top of SMT~\cite{SMTLib2010,z3} by discriminating over the function-value variants flowing into the call site: if $f$ is known, the invocation is replaced by the effect of the concrete function; otherwise (e.g., a function-typed parameter or a closure loaded from storage), the behavioral predicates of $f$ describe its effect.

    \item We extend \MVP's \emph{specification inference} tool \cite{GrieskampEtAl26Inf} to function-valued code.
    The tool computes weakest preconditions over Move bytecode with closures and dynamic dispatch, automatically producing behavioral-predicate-based specifications.
    This both reduces the annotation burden and yields a validation pipeline confirming the encoding is internally consistent.
\end{Itemize}

\noindent While these aspects have similarities to features and solutions in Dafny, \fstar, or \verus, they are novel in the details. For example, Dafny has state labels but does not support higher-order functions with side effects, nor quantification over state labels. See Sec.~\ref{sec:related} for a more thorough discussion of related work.



\Section{Using Function Values in Move Specifications}
\label{sec:move-hof}

This section recalls the aspects of Move on Aptos \cite{MoveOnAptosBook} that matter for the verification of higher-order programs, introduces an automated market maker (AMM) as a running example, and illustrates the common specification patterns our extension supports.
All examples can be also found at \cite{PaperExamples}.

\SubSection{Move on Aptos} \label{sec:move}

Move is a bytecode-verifiable programming language designed from the outset to coexist with formal verification.
For the purposes of this paper, the relevant properties are:

\begin{Itemize}
    \item \emph{Strong static typing with generics.}  Every expression has a statically known explicit type; generic functions and structs are monomorphized for verification~\cite{DillGPQXZ22}.

    \item \emph{Type-indexed global storage.}  A struct type with the !key! ability is called a \emph{resource} and can be stored in global state, which is a map keyed by (resource type, address) pairs: a resource can be moved to, borrowed, and moved from the location !T[addr]!, with linear ownership enforced statically, and !exists<T>(addr)! tests its presence.  Specifications quantify over this memory directly, and !modifies! clauses name the footprint a function may touch, providing a frame condition.

    \item \emph{Deterministic execution.}  A Move function's results and effects are fully determined by its arguments and the pre-state of global storage; there is no in-language concurrency or other source of non-determinism.

    \item \emph{Reference semantics.}  References are borrowed from local variables or from global storage; this yields an alias-free memory model that makes verification tractable~\cite{DillGPQXZ22}.

    \item \emph{Strongly-typed bytecode.}  Move source is compiled to bytecode whose type- and memory-safety are checked before execution.  \MVP verifies this bytecode against specifications in the source.

    \item \emph{Integrated specification language.}  Move lets developers write pre-/postconditions, abort conditions, data invariants, and global memory invariants in the same source file as the code they describe, or optionally,
    into separate files.
\end{Itemize}

\Paragraph{The Move Prover}
\MVP is an automated formal verifier built into the Move toolchain;
the version discussed here and described in~\cite{DillGPQXZ22} is a complete rewrite of the earlier version in~\cite{MOVE_PROVER}.
Developers annotate Move source with specifications --- preconditions, postconditions, abort conditions, data invariants, and loop invariants --- using a first-order specification language that shares Move's type system.
\MVP translates the annotated program to Boogie~\cite{boogie,thisisboogie2}, an intermediate verification language, which Z3~\cite{z3} then discharges as SMT queries.
The tool targets routine use during development: verification is designed to complete within seconds per function through careful SMT encoding and monomorphized, modular function summaries; it is used in CI (continuous integration testing) in the workflows at Aptos.
Abort conditions deserve special note: multiple !aborts_if! clauses form a \emph{biconditional} --- the function aborts if and only if at least one clause holds --- rather than the one-directional ``if it aborts, then ...'' familiar from partial-correctness Hoare logic.

\Paragraph{Function values in Move}
A new version of Move extends the type system with \emph{function types} $T_1 \times \dots \times T_n \to T$ (written !|T1,...,Tn|T! in the concrete syntax) and allows values of such types to be constructed from concrete function symbols or from lambda expressions, possibly with captured arguments.
In Move's type ability system, function values may be marked !copy! (may be duplicated), !drop! (may be discarded), and !store! (may be placed in struct fields and thereby persisted to global storage).
References may not be captured; only plain values may appear in a closure's payload, ensuring that captured data is independent of any borrow scope.

\SubSection{Automated Market Maker Example}
\label{sec:amm}

The running example throughout this paper is a simplified AMM (Automated Market Maker) module in which the \emph{pricing curve} is a function value stored in the AMM pool.
An AMM holds reserves of two assets in a pool and lets traders exchange one for the other, the price determined by a curve over the current reserves rather than by an order book; prominent protocols include Uniswap~\cite{UniswapV2}, Curve, and Balancer.
The pricing function has to satisfy a number of invariants: (i)~the function never aborts, (ii)~the output never exceeds the output reserve, (iii)~the output is monotone in the input amount, and (iv)~the product of the two reserves does not decrease after a swap (constant-product preservation).
The spec expresses these invariants using \emph{behavioral predicates} and \emph{state labels}, two constructs introduced by this paper:

\begin{MoveBox}
  struct Pool has key {
    reserve_x: u64, reserve_y: u64,
    // (reserve_in, reserve_out, amt) = out
    pricing: |u64, u64, u64|u64 has copy, store
  }
  spec Pool {
    // Function allowed to read any global state.
    reads_of<self.pricing> *;

    // Pricing must never abort.
    invariant forall S in *, ri: u64, ro: u64, a: u64:
      S |~ !aborts_of<self.pricing>(ri, ro, a);

    // Output never exceeds the output reserve.
    invariant forall S in *, ri: u64, ro: u64, a: u64:
     S.. |~ result_of<self.pricing>(ri, ro, a) <= ro;

    // Monotonicity in the input amount.
    invariant forall S in *, ri: u64, ro: u64,
                        a1: u64, a2: u64:
      S.. |~ a1 <= a2 ==>
        result_of<self.pricing>(ri, ro, a1)
             <= result_of<self.pricing>(ri, ro, a2);

    // Constant-product preservation.
    invariant forall S in *, ri: u64, ro: u64, a: u64:
      S.. |~
        (ri + a)
        * (ro - result_of<self.pricing>(ri, ro, a))
          >= ri * ro;
  }
\end{MoveBox}

Each invariant is universally quantified over a state label !S!, ranging over all states reachable during execution in which the pool exists at some address.
A \emph{state label} such as !S! names a program point; the state at !S! comprises both the global memory state and the local state (via references) at that point.
For the Pool invariants the local component is not relevant, since !pricing! takes only plain value arguments and no references; the global component covers the resources declared by !reads_of<self.pricing> *!.

State labels appear in two forms.
A \emph{single-state} predicate $\slabeled{S}{p}$ evaluates $p$ entirely in state $S$; the no-abort invariant uses this form because whether !pricing! aborts depends only on the pre-state of the call.
A \emph{two-state} predicate $\slabeled{S..}{q}$ takes $S$ as the pre-state of the !pricing! invocation and the resulting post-state as the second end; the output, monotonicity, and constant-product invariants use this form because the return value of !pricing! is only defined after execution from $S$.
More generally, $\slabeled{S_1..S_2}{q}$ names both ends explicitly; in the example, $\slabeled{S..}{q}$ fixes $S$ as the pre-state, with the post state unspecified.

The behavioral predicates !aborts_of!, !result_of!, and (elsewhere) !ensures_of! and !requires_of!, take a function value and its arguments and evaluate to the corresponding aspect of that call's behavior.
Without behavioral predicates, specifying a higher-order function would require inlining the callee's conditions at every call site, leading to a combinatorial explosion in specification size; behavioral predicates make reasoning about function values modular and abstract.
Using them in invariants as in the example creates (i)~a requirement for any code creating a pool to pass only compliant function values, and (ii)~a guarantee for any code calling a pool function that the requirements are met. Both conditions are verified by \MVP.

Notice that the monotonicity invariant relates two invocations of the same function value --- a \emph{relational} property (hyperproperty).
This is no special case: per state pair, !result_of! is a mathematical function of the function value and its arguments, so any first-order formula relating multiple !result_of! terms is a legal specification.
For example, !S.. |~ result_of<f>(x) > result_of<g>(x)! compares two function values pointwise; relations across different states are expressed via distinct state labels.

\Paragraph{Swapping}
The major functionality of the pool is to facilitate !swap! (exchange) of one asset with another.

\begin{MoveBox}
  fun swap(pool: &mut Pool, amt: u64): u64 {
    let out = (pool.pricing)
              (pool.reserve_x, pool.reserve_y, amt);
    pool.reserve_x += amt; pool.reserve_y -= out;
    out
  }
  spec swap {
    // Can only abort on reserve overflow.
    aborts_if pool.reserve_x + amt > MAX_U64;
    // Outcome is determined by pricing function.
    ensures result == old(result_of<pool.pricing>(
              pool.reserve_x, pool.reserve_y, amt));
    // Preserves product in pool.
    ensures pool.reserve_x * pool.reserve_y
         >= old(pool.reserve_x * pool.reserve_y);
  }
\end{MoveBox}

\noindent In a postcondition, !result! is the predefined name of the function's return value, and !old(e)! evaluates !e! in the function's pre-state.
The specification of !swap! can be verified by \MVP because of the invariants guaranteed for the (dynamic) !pricing! function. In this case, we only illustrate the aborts condition and product preservation. For aborts, !pricing! itself is non-aborting, and (more subtly), |pool.reserve_y - out| cannot underflow because of the 'output never exceeds the output reserve' invariant, so only the overflow condition remains. (Notice \MVP would flag any aborts conditions not covered). For product preservation, the constraint can be derived as well.

\Paragraph{Pricing variants}
At the time a pool value is constructed, the invariants for the pricing function need to be verified.
Obviously this is more difficult than the application side, since now properties of functions need to be proven.
In reality, those proofs will often need custom solutions, but for the given example, \MVP can handle some cases gracefully.
First, we define a function which computes the so-called \emph{constant product}, which is a pricing function describing a direct swap without fees; however, rounding errors can lead to a loss in returned assets:

\begin{MoveBox}
  fun product(ri: u64, ro: u64, a: u64): u64 {
    let num = (ro as u128) * a;
    let den = (ri as u128) + a;
    if (den == 0) 0 else (num / den) as u64
  }
\end{MoveBox}

\noindent When constructing a pool using !Pool{.., pricing: product}!, \MVP is able to discharge all verification conditions.

The next example uses a pricing function which charges a configurable fee:

\begin{MoveBox}
  struct Fee has key {bps: u64}
  fun with_fee(owner: address,
               ri: u64, ro: u64, a: u64): u64 {
    let b = (a as u128)
       * (10000 - Fee[owner].bps as u128) / 10000;
    product(ri, ro, b as u64)
  }
  spec with_fee {
    aborts_if !exists<Fee>(owner);
    aborts_if Fee[owner].bps > 10000;
    ..
  }
\end{MoveBox}

\noindent In order to construct a pricing function we need to curry the address for the fee resource and capture it in a closure, which can be done in Move as below:

\begin{MoveBox}
    let owner: address;
    Pool{
      ..,
      pricing: |ri, ro, a|
        with_fee(owner, ri, ro, a)
    }
\end{MoveBox}

\noindent \MVP detects that !with_fee! violates the requirement to not abort.
Guarding the definitions with checks and using a default fee fixes this and makes verification succeed.

\begin{MoveBox}
  fun with_fee_compliant(owner: address,
               ri: u64, ro: u64, a: u64): u64 {
    let bps =
      if (!exists<Fee>(owner)
          || Fee[owner].bps > 10000)
        500
      else
        Fee[owner].bps;
    let b =
        (a as u128) * (10000 - bps as u128) / 10000;
    product(ri, ro, b as u64)
  }
  spec with_fee_compliant {
    aborts_if false;
    ..
  }
\end{MoveBox}

Notice that currently, a user needs to provide complete aborts conditions of even simple functions as |with_fee|.
While \MVP in general allows to keep aborts conditions unspecified this does not work right now for aborts conditions in the case of function values.
We are looking at connecting specification inference in \MVP to the use case of compliance for function values; see also Sec.~\ref{sec:inference}.

The AMM example shows that non-trivial invariants can be specified for function values, and that the prover is capable of verifying them on the construction and caller sides, even though the invariants presented constitute non-linear arithmetic problems. While determining that the aborts contract is violated by |with_fee| appears easy (after all, one can directly see in the spec that the function aborts), there are more subtle things happening in the background -- for example, that |product| does not have any underflow on subtraction because of the |pricing| invariant that the return value does not exceed the reserve.

However, we have no illusions about the scalability of compliance-side verification.
In general, passing a function $f$ as an actual parameter $p$ requires proving $\forall x: \requiresof{p}{x} \implies \requiresof{f}{x}$, $\forall x: \abortsof{f}{x} \iff \abortsof{p}{x}$, and $\forall x,y: \ensuresof{f}{x, y} \implies \ensuresof{p}{x, y}$; Sec.~\ref{sec:enc-bp} describes how these obligations materialize in the encoding.
For the examples in this paper, !product! and the application side (!swap!, the callers of !find!) verify from the shown specifications alone; the nonlinear fee variants need in addition a one- or two-line proof hint (a case split, and a stepping-stone assertion bounding the fee-adjusted input).
For harder universal properties of functions, we do not expect compliance proofs to be automated easily; rather, they will require more substantial proof hints and/or trusted assumptions.
Nevertheless, on the application side of function values, such as inside the |swap| function, verification is not more complex than for first-order functions.

\SubSection{Filter-style Loops} \label{sec:find}

Behavioral predicates are not limited to struct invariants; they appear naturally in loop invariants and function specs for higher-order collection operations.
As a minimal illustration, consider a generic !find! which returns the first index of !v! whose element satisfies a predicate !pred!, or !len(v)! when no such element exists.
A spec-level helper !no_match_before! captures the recurring ``no earlier element has matched'' pattern:

\begin{MoveBox}
   fun find<T>(v: &vector<T>,
              pred: |&T|bool has copy + drop): u64 {
    let i = 0;
    let n = v.length();
    while (i < n) {
      if (pred(v[i])) return i;
      i += 1;
    } spec {
      invariant i <= n;
      invariant no_match_before(v, pred, i);
    };
    n
  }
  spec find {
    // opaque: caller side uses spec only
    pragma opaque;
    ensures result <= len(v);
    ensures no_match_before(v, pred, result);
    ensures result < len(v)
              ==> result_of<pred>(v[result])
  }
  spec fun no_match_before<T>(v: vector<T>,
                 pred: |&T|bool, end: u64): bool {
    forall j in 0..end: !result_of<pred>(v[j])
  }
\end{MoveBox}

\noindent !no_match_before! is an ordinary spec function, but its body calls the behavioral predicate !result_of<pred>! on each element --- a predicate-parametric abstraction is thus first-class in Move and can be reused across loop invariants and post-conditions.
The function's post-condition characterizes !result! entirely through the same evaluator: either !result! is a valid index whose call is the first to match, or !result == len(v)! and no element matches at all.
No concrete predicate is ever named in either the body or the spec; a caller instantiates !pred! with any function value, and the specification is discharged through its behavioral contract.

To illustrate the proof obligation at the caller, consider specializing !find! with an inline lambda whose behavioral contract is given by an inline !spec! clause.
As was discussed earlier for |with_fee|, functions provided as parameters need to have explicit specifications, therefore the lambda has a trivial spec block:

\begin{MoveBox}
  fun find_zero_lambda(v: &vector<u64>): u64 {
    find(v, |x| x == 0
             spec { ensures result == (x == 0) })
  }
  spec find_zero_lambda {
    ensures result <= len(v);
    ensures result < len(v) ==>
      v[result] == 0 &&
      (forall j in 0..result: v[j] != 0);
    ensures result == len(v) ==>
      (forall j in 0..len(v): v[j] != 0);
  }
\end{MoveBox}

\noindent Since !find! is opaque, \MVP reasons from !find!'s spec alone. (Without opaque, \MVP uses spec and body combined at caller side.)
At the match index, !find! guarantees $\resultof{pred}{v[result]}$, which --- by the lambda's spec --- reduces to $v[result] = 0$.
For earlier indices, !no_match_before! constrains $\resultof{pred}{v[j]}$ to be false for every index $j$ before $result$, which reduces to $v[j] \neq 0$.
\MVP discharges all three !ensures! clauses of !find_zero_lambda! without ever inspecting the lambda body.

Effectively, with the higher-order function |find|, we have verified a summary of this specific loop pattern which can now be reused many times on the caller side. Users calling functions like |find|, |map|, |reduce|, etc. will hardly need to write loop invariants themselves.

\SubSection{More About State Labels}
\label{sec:state-labels}

The Pool invariants in Sec.~\ref{sec:amm} used \emph{universally} quantified state labels: every state in which the pool exists must satisfy the pricing invariants.
State labels have a second, equally important use: \emph{existentially} quantifying over an intermediate state to express that a computation proceeds through a specific intermediate point.
Consider a higher-order function that sequences two stateful operations on a shared value:

\begin{MoveBox}
  fun followed_by(f: |&mut A|, g: |&mut A|, x: &mut A) {
    f(x); g(x)
  }
  spec followed_by {
    reads_of<f> *; reads_of<g> *;
    ensures exists S in *:
      (..S |~ ensures_of<f>(x)) &&
      (S.. |~ ensures_of<g>(x))
  }
\end{MoveBox}

\noindent The existentially quantified label !S! witnesses the intermediate state between the two calls.
Since no !modifies_of! is declared for !f! or !g!, they are assumed to have no global memory modification; the local component of the state at !S! is therefore the updated value of !x!, and the global component is unchanged.
The conjunct $\slabeled{..S}{\ensuresof{f}{x}}$ asserts that the postcondition of |f| holds from the pre-state of !followed_by! up to !S!; the conjunct $\slabeled{S..}{\ensuresof{g}{x}}$ asserts that the postcondition of !g! holds from !S! to the final post-state.
Crucially, the post-state of !f! automatically becomes the pre-state of !g!, without any explicit threading of intermediate values through the specification.

State labels are not restricted to higher-order functions.
When !f! and !g! are concrete named functions, the same pattern allows a wrapper to be specified by composing the behavioral predicates of its callees, avoiding duplication of their postconditions.


\Section{Encoding}
\label{sec:enc}

We describe the implementation of function values as a translation into a conceptual (pseudo-code style) imperative intermediate verification language (IVL) inspired by Boogie~\cite{boogieIVL,thisisboogie2}.
The encoding is given at the IVL level because this is where the interesting design decisions become visible --- the subsequent step to pure SMT is straightforward and uses standard techniques.
The fragment of the IVL used in this section is familiar: procedures with $\HAVOC$, $\ASSUME$, $\ASSERT$, pure functions, \emph{algebraic datatypes} (tagged unions with selectors, like Rust enums), universal \emph{axioms} with \emph{triggers}, and polymorphic map types.
State is represented by Boogie-style maps; in particular, each resource type $\tau$ yields a map $M_\tau \in \Addr \rightarrow \tau$, with $\Addr$ being the domain of addresses in Move (256-bit integers).
The representation is similar to that in the earlier \MVP paper~\cite{DillGPQXZ22}.

In the following we focus on the aspects specific to function values: the representation of function values (Sec.~\ref{sec:enc-fvals}), the memory access and modification sets used to define the invocation schema (Sec.~\ref{sec:enc-mem}), the encoding of invocation of function values (Sec.~\ref{sec:enc-invoke}), the encoding of behavioral predicates via axioms (Sec.~\ref{sec:enc-bp}), and finally the encoding of intermediate state labels (Sec.~\ref{sec:enc-frame}).
Everything in this section operates \emph{after} monomorphization, so type parameters have been specialized and the set of closure, parameter, and field variants reaching a given function type is known statically~\cite{DillGPQXZ22}.

\SubSection{Representing Function Values}
\label{sec:enc-fvals}
For each function type $\tau = (T_1,\ldots,T_n) \to T$ that occurs in the monomorphized program, we create a datatype in the IVL which represents all the potential \emph{sources} of function values in a given program.
To this end, \MVP's mono-analysis records three sets: the set $\Clo_\tau$ of concrete functions $f$ constructing closures of type $\tau$, the set $\Par_\tau$ of function-typed parameters of verification targets, and the set $\Fld_\tau$ of storable function-typed struct fields whose type is $\tau$.
The union of these three sets is materialized as an algebraic datatype in the IVL --- one constructor per variant:

\begin{ivl}
\DATATYPE \tau\ \{\\
\t1 C_{f}(\sq{c : T}), \ldots, P_{f,x}(), \ldots, F_{\sigma.x}(n : \nat), \ldots \\
\}
\end{ivl}

\noindent Here $\sq{c:T}$ are the types of the values captured by closure $f$; $P_{f,x}$ is the parameter variant for function-value parameter $x$ of function $f$; and $F_{\sigma.x}$ is the field variant for struct field $x$ of struct type $\sigma$ (we use $\sigma$ here to avoid confusion with the state-label variable $S$ introduced in Sec.~\ref{sec:enc-frame}).
Parameter variants are nullary because, in a well-typed verification condition (VC), the value of an entrypoint function-typed parameter is a skolem constant pinned to its variant at entry; no runtime payload is required.
Field variants carry an integer discriminator $n$ so that two locations in memory holding distinct field values are not forced to be equal --- a datatype with a single nullary constructor would otherwise collapse all field values to one.

Notice that this representation captures \emph{all} possible functions of type $\tau$ relevant for a given VC.
So we do not actually need to deal with an arbitrary number of function values when it comes to higher-order function verification, but only those which can be derived from the program in a given context.
The representation is further narrowed if the function value of a given type is not stored in a structure which ends in global memory.
Storing a function value in a struct field introduces dynamic dispatch: two different field locations may hold different closures, and the discriminator $n$ in $F_{\sigma.x}(n)$ ensures the solver treats them as distinct.
Nevertheless, we are able to pinpoint a specification to a particular $F_{\sigma.x}(n)$, as illustrated by the AMM example (Sec.~\ref{sec:amm}).

\SubSection{Computing Memory Accesses}
\label{sec:enc-mem}

Invoking a function value requires reasoning about which memory maps a given function type $\tau$ may read ($\Mem_\tau$) and which locations it may modify ($\Mod_\tau(\sq{p})$); we define both here before presenting the invocation schema.
Intuitively, the access set is the \emph{read footprint} of the function values of type $\tau$ --- it delimits which parts of memory their behavior may depend on --- while the modification set is their \emph{write footprint}, providing the frame condition: everything outside of it is unchanged by an invocation.
To recall, in the Move language, persistent global storage, or memory, can be accessed via type indexing.
For instance, if !R! is a resource type, we write !&mut R[addr]! in the language to obtain a mutable reference to the memory which can then be subsequently mutated.

\MVP's representation of memory uses type-indexed maps $M_\rho \in \Addr \fun \rho$, with $\rho$ a resource type (a !struct! in Move with the !key! ability).
In the imperative procedures of the IVL, those maps are implicitly available as global variables.
But in pure expressions like behavioral predicates, they need to be made explicit as parameters; this is why $\Mem_\tau$ appears as an explicit argument to behavioral predicates throughout this section.

A few definitions for memory maps are needed.
Note that these are definitions on the meta level of the description here and not part of the IVL; they are used to describe pseudo code in terms of the IVL.

First, with $\Mem_R = \{ M \in \Addr \fun \rho \mid \rho \in R\}$ we denote a set of memory maps for the resource types in $R$.
The $R$ can be omitted if it is clear from the context. For convenience, we write $\Type(\Mem)$ for the set of all types $\rho$ used in the memory maps of $\Mem$.
With $\Mem_R \downarrow S$, where $S \subseteq R$, we denote the projection $\{ M \in \Mem_R \mid \rho(M) \in S \}$.

In addition to memory, we also need to reason about modifications.
A modification is a set of pairs of a type and an expression of the IVL of type $\Addr$, $\Mod \in \power (\Type \cross \Expr)$.
We write $\Mod(\sq{p})$ for a set of modifications built on expressions referencing parameters in $\sq{p}$.
For instance, the declaration |modifies_of<f>(a: address) R[a]| (introduced below) contributes the pair $(R, a)$; at an invocation of !f!, $a$ is bound to the actual argument, so $\Mod(\sq{p})$ gives the modified locations as a function of the arguments $\sq{p}$.

\Paragraph{General Access}
Memory access $\Mem_\tau$ is computed as the union of all the accesses in the variants of $\tau$, $\Mem_{C_f} \cup \Mem_{P_{f,x}} \cup \Mem_{F_{\sigma.x}}$.
For a closure, $\Mem_{C_f}$  can be statically derived from the code.
For parameters and fields, the information is declared in the spec block:

\begin{MoveBox}
  spec foo(f: |T|S, g: |T|S) {
    reads_of<f> R; // f can read resource R
    reads_of<g> *; // g can read any resource
                   // in scope of the VC
  }
\end{MoveBox}

\noindent For the wildcard operator !*!, all memory which is read or written as part of the currently generated VC is included, plus an abstract memory map for any `unknown' memory the function may depend on.
The unknown memory is needed for a sound encoding.
For instance, if we verify function !foo! above in a context where no memory is involved anywhere, we must not assume that !g(x)! delivers the same value in different states, as it may depend on this unknown additional memory.
However, we can still know that the specification of !g! holds in every possible state.

\Paragraph{Write Accesses}
Similarly to general access, the modifies set $\Mod_\tau$ is computed as the union of all the modifications in the variants of $\tau$, $\Mod_{C_f} \cup \Mod_{P_{f,x}} \cup \Mod_{F_{\sigma.x}}$.

For closures, $\Mod_{C_f}$ is derived from the code and, if present, the !modifies R[a]! declaration in the language.
These existed before higher-order functions, for specifying what functions modify.

For parameters and fields, the information is declared in the spec block, where for the !modifies_of<f>! clause, the parameters of the function value itself are passed:

\begin{MoveBox}
  spec foo(f: |address|) {
    // f can modify resource R[a]
    modifies_of<f>(a: address) R[a];
  }
\end{MoveBox}

\noindent If no !modifies_of! is declared, it is assumed that the function does not modify global memory.

Notice that a wildcard (!modifies_of<f> *!) is not supported, since the invocation of !f! would render the memory arbitrary (no effective frame condition).
Consider a generic utility !for_each_account(f: |address|)! applying a caller-supplied, state-mutating callback: one would like to specify it once, for callbacks with open-ended write footprints, but today the footprint must be declared per resource, as in !modifies_of<f>(a) R[a]!.
Lifting this restriction is left for future work.

\SubSection{Invoking Function Values}
\label{sec:enc-invoke}

When a function value is invoked, we are looking at either a concrete closure value $C_{g}$ or a parameter or field selection as described above.
In the former case, one can simply delegate to calling the actual underlying function $f$, by composing the captured and provided arguments into a full argument list.

In the other cases of parameters and field selections, we use the behavioral predicates as defined for the function value, using the standard schema for encoding a call to a Move function via the function's specification.
This reduces the problem of invocation to how those behavioral predicates are encoded.
In the procedure below, $\Mem_\tau$ denotes the projection of the current global memory state to the resource types in $\tau$'s access set (as computed in Sec.~\ref{sec:enc-mem}).
\begin{ivl}
\PROC invoke_\tau(x: \tau, \sq{p:T}) \RETURNS (\sq{r: T}) \\
\t1 \IF x \IS\ C_{f}(\sq{c}) \\
\t2    \sq{r} := \CALL f(\sq{c}, \sq{p}) \\
\t1 \ELSE \\
\t2    \ASSERT requires_\tau(x, \Mem_\tau, \sq{p}) \\
\t2    \IF\ aborts_\tau(x, \Mem_\tau, \sq{p}) \\
\t3       abort\_flag := \TRUE \\
\t2    \ELSE \\
\t3       \LET \Mem'_\tau := \Mem_\tau \\
\t3       \HAVOC \Mod_\tau(\sq{p}),\ \bar{r} \\
\t3       \ASSUME ensures_\tau(x, \Mem'_\tau, \Mem_\tau, \sq{p}, \sq{r}) \\
\end{ivl}

\noindent Here $\LET \Mem'_\tau := \Mem_\tau$ freezes the pre-invocation memory into $\Mem'_\tau$, while $\Mem_\tau$ becomes the post-invocation state after $\HAVOC$.
We therefore pass $\Mem'_\tau$ as the pre-state and $\Mem_\tau$ as the post-state to $ensures_\tau$ (note that $\Mem'$ consistently denotes the pre-state throughout this section).
$\HAVOC \Mod_\tau(\sq{p})$ is shorthand for havocing, for each $(\rho, a) \in \Mod_\tau(\sq{p})$, the entry $M_\rho[a]$ of the type-indexed memory map for $\rho$.
$abort\_flag$ is a prover-internal boolean global that signals an abort to the enclosing verification condition, consistent with how \MVP models Move's abort semantics throughout the IVL.
The $\IF$ branch reduces to a standard function call $\CALL f(\sq{c}, \sq{p})$, which \MVP encodes using the same $\ASSERT$/$\HAVOC$/$\ASSUME$ schema as the $\ELSE$ branch, specialized to $f$'s declared spec.
When $f$ is opaque, the two branches are in fact semantically equivalent: an opaque call is already compiled to this schema driven by $f$'s spec, matching what the $\ELSE$ branch does for the abstract variant.
The $\ELSE$ branch therefore introduces no new encoding machinery; it applies the existing opaque-call schema, driven by the behavioral predicates of the unknown variant $x$ rather than a concrete function's spec.

\SubSection{Behavioral Predicates}
\label{sec:enc-bp}

Behavioral predicates are derived by switching over the variants of the function value representation $\tau$, similarly to the $invoke_\tau$ procedure above, delegating extraction of the underlying predicates to per-variant functions.
Intuitively, each behavioral predicate is total over the variants of $\tau$: applied to a closure, it unfolds to the underlying function's specification; applied to an abstract parameter or field variant, it remains an opaque symbol of which only the user-declared conditions are known.

Below, the definition for ensures is given.
The other predicates are defined similarly.
Specific to ensures is that it needs to take care of the frame condition: the caller havoced the full modification set, but only a subset of that is constrained by the given variant.
The handling needs to account for the fact that modification addresses are symbolic expressions that may alias.

In what follows, we write $\mathit{pred}_\tau[\Mem', \Mem](\ldots)$ with square brackets for the memory pair: $\Mem'$ is the pre-state and $\Mem$ the post-state, following the convention established in Sec.~\ref{sec:enc-invoke}.
These memory arguments must be passed explicitly to pure functions, whereas they are implicit globals in IVL procedures.
We write $\Mod_x$ for the modification set of the specific variant of $x$ --- that is, $\Mod_{C_f}$ when $x = C_f(\sq{c})$, $\Mod_{P_{g,y}}$ when $x = P_{g,y}()$, and $\Mod_{F_{\sigma.x}}$ when $x = F_{\sigma.x}(n)$.
We write $\Mem'(\rho)$ for the map $M_\rho$ in the pre-state, and $\Mem(\rho)$ for the map $M_\rho$ in the post-state.

\begin{ivl}
  \FUN ensures_\tau[\Mem', \Mem](x: \tau, \sq{p: T}, \sq{r: U}): \BOOL \\
\t1  \bigwedge_{(\rho, a) \in \Mod_\tau}( \\
\t2    \IF \rho \notin \Type(\Mod_x) \\
\t3       \Mem(\rho) = \Mem'(\rho) \\
\t2    \ELSE \\
\t3       (\bigwedge_{(\rho, a') \in \Mod_x} a' \neq a) \\
\t4          \implies \Mem(\rho)[a] = \Mem'(\rho)[a] \\
\t1  ) \land ensures_x[\Mem'_x, \Mem_x](p, r)
\end {ivl}

The relational representation of a function via ensures does not yet capture that Move functions are \emph{deterministic} (Sec.~\ref{sec:move}): for a fixed function value, arguments, and pre-state, there is exactly one outcome.
Functional behavior is established by introducing $result_\tau$ as an uninterpreted function, connected to ensures via an axiom (from here on, quantifiers are annotated with their \emph{trigger} in braces, further discussed in Sec.~\ref{sec:imp-notes}):
\begin{ivl}
  \FUN result_\tau[\Mem', \Mem](x: \tau, \sq{p: T}) \fun \sq{U} \\
  \AXIOM \forall x, \sq{p} @ \{result_\tau[\Mem', \Mem](x, \sq{p})\} \\
\t1 \LET \sq{r} := result_\tau[\Mem', \Mem](x, \sq{p}) @ \\
\t2 ensures_\tau[\Mem', \Mem](x, \sq{p}, \sq{r})
\end {ivl}

The result tuple $\sq{U}$ comprises the declared results of $\tau$ and, per mutable-reference parameter, the post-state of the referenced value; thus both the explicit results and the implicitly modified local state are pinned as a function of the function value, its arguments, and memory.
For a resource which the variants of $\tau$ only read, a single memory map is passed, so the outcome is determined by the pre-state alone, as determinism demands; the post-state of \emph{modified} global memory remains relationally constrained by the invocation schema (havoc within the frame $\Mod_\tau$, then assuming ensures).

The axiom is deliberately one-directional, stating that the tuple chosen by $result_\tau$ satisfies $ensures_\tau$.
The seemingly natural biconditional $ensures_\tau(x, \sq{p}, \sq{r}) \iff \sq{r} = result_\tau(x, \sq{p})$ would force \emph{all} solutions of ensures to coincide --- inconsistent as soon as the declared ensures of some variant under-constrains its outcome, such as a mere |ensures result > 0|.

The actual conditions extracted depend on the variant.
For closures, they are directly extracted from the spec block.
For parameters and fields, we introduce uninterpreted functions; for ensures specifically, we introduce an uninterpreted per-variant result function and \emph{define} the variant's ensures as its graph, so functional behavior holds by construction (the field variant is analogous to the parameter variant shown, with the discriminator $n$ as additional argument):

\begin{ivl}
  \FUN ensures_{C_f(\sq{c})}[\Mem', \Mem](\sq{p: T}, \sq{r: U}): \BOOL \\
\t1    \langle \textit{extract from spec block of}~~f(\sq{c}, \sq{p}) \rangle \\
  \FUN result_{P_{f,x}}[\Mem', \Mem](\sq{p: T}) \fun \sq{U} \\
  \FUN ensures_{P_{f,x}}[\Mem', \Mem](\sq{p: T}, \sq{r: U}): \BOOL \\
\t1    \sq{r} = result_{P_{f,x}}[\Mem', \Mem](\sq{p}) \\
\end{ivl}

\noindent The meaning of the uninterpreted functions is determined by how they are injected into a VC via the regular mechanism by which pre-/postconditions and invariants are handled.
Consider a precondition like |requires !aborts_of<param>(x)|, where !param! is a function-value parameter of a function !foo!.
This will be injected in the VC as $\ASSUME \lnot aborts_{P_{foo, param}}(x)$.
Similarly, for a data invariant of a field, an assumption will be generated when the field is selected.

\Paragraph{Compliance checking}
The same mechanism yields the compliance obligations of Sec.~\ref{sec:amm} when a concrete function value flows into an abstract position.
Since behavioral predicates are first-order, compliance verification mimics the conventional handling of pre-/postconditions: the condition |requires !aborts_of<param>(x)| above, assumed on the definition side of !foo!, is \emph{asserted} at call sites of !foo!, where !param! is bound to an actual function value --- say a closure $C_g(\sq{c})$ --- so the assertion reduces, via the per-variant extraction, to a condition over $g$'s specification.
Data invariants over function-valued fields, as in the Pool example, create the same obligations at pack time; constraints via $result_\tau$, including relational ones, are discharged through the result axiom.
The resulting VCs quantify over the function value's arguments (and possibly memories) and may involve non-linear arithmetic (Sec.~\ref{sec:amm}); they are inherently harder than application-side VCs and may need proof hints.

\SubSection{Intermediate State Labels}
\label{sec:enc-frame}

Recall from Sec.~\ref{sec:move-hof} that a state label !S! denotes a program point.
A specification mentioning !S! talks about a state \emph{in the middle} of a function's execution, invisible in a pre-/post-state contract; the encoding materializes it as an existentially quantified snapshot, connected to both ends by the conditions mentioning !S!.
Formally, a \emph{state} at label $S$ is a pair $(\Mem^S, \bar{v}^S)$ consisting of a memory snapshot and a tuple of local values.
The memory component $\Mem^S$ has the same type as the ambient memory: a tuple of type-indexed maps $M_\rho \in \Addr \to \rho$ for each resource type $\rho$ in scope at that program point, quantified existentially in the IVL.
The local component $\bar{v}^S$ captures values of mutable-reference parameters at the intermediate point: for a closure over a mutable reference of type $T$ with no global memory effects, $\bar{v}^S$ is a single value $x^S : T$, and $\Mem^S$ is vacuous (the function touches no global resources).
When a closure both accesses global memory and takes mutable references, both components are non-trivial.
Here we focus on the global-memory case (vacuous $\bar{v}^S$); Sec.~\ref{sec:imp-notes} notes how the mutable-reference case is handled.

Consider the following specification using intermediate state labels:

\begin{MoveBox}
  struct R(u64) has key;
  spec foo(f: |address|, g: |address|, a: address) {
    ensures ..S |~ ensures_of<f>(a)
    ensures S |~ R[a].0 > 0;
    aborts_if S |~ aborts_of<g>(a);
    ensures S.. |~ ensures_of<g>(a)
  }
\end{MoveBox}

As outlined in the last section, we want to be able to extract behavioral predicate functions from this specification.
For the ensures we can come up with the following, which simply replicates part of the spec:

\begin{ivl}
  \FUN ensures_{C_{foo}}[\Mem', \Mem](a) \\
\t1  \exists \Mem^S @
      \<ensures_\tau(f,\Mem',\Mem^S,a) \land R[\Mem^S][a].0 > 0 \land \\
        ensures_{\tau'}(g,\Mem^S,\Mem,a)\>
\end{ivl}

For the aborts which happens at the intermediate state !S!, the state !S! needs to be first established before the aborts condition can be checked for !g!.
To achieve this, all ensures conditions which reach !S! are combined with the aborts condition, as in:

\begin{ivl}
  \FUN aborts_{C_{foo}}[\Mem', \Mem](a) \\
\t1  \exists \Mem^S @
      \<ensures_\tau(f,\Mem',\Mem^S,a) \land R[\Mem^S][a].0 > 0 \land \\
        aborts_{\tau'}(g,\Mem^S,a)\>
\end{ivl}

This is semantically sound because we can assume inductively that the function $foo$ has successfully verified. For a successfully verified function, where $\sq{A}$ are the |aborts_if| conditions and $\sq{P}$ the |ensures| conditions, it holds:
$$
   \bigvee \sq{A} \lor \bigwedge \sq{P}
$$

\noindent That is, either the function aborts under one of the given conditions, or all of its ensures conditions hold.

In general, since the relation between state labels spawned by predicates of the form $\slabeled{S_1..S_2}{p}$ must be acyclic, one can extract from $\sq{P}$ those conditions which lead into $\Mem^S$, and from $\sq{A}$ those which start from state $\Mem^S$.
Given this, the aborts condition can be synthesized using the following scheme:
$$
    \bigwedge \sq{P}_{..\Mem^S} \land (\bigvee \sq{A}_{\Mem^S} \lor
    \bigwedge \sq{P}_{\Mem^S..\Mem^T} \land (\bigvee \sq{A}_{\Mem^T} \lor
    \bigwedge \ldots ))
$$

\noindent Here, $\sq{P}_{..\Mem^S}$ establish state $\Mem^S$ in which we check for the aborts condition $\sq{A}_{\Mem^S}$,
or recurse over the remaining $\sq{P}$ and $\sq{A}$.

\SubSection{Implementation Notes}
\label{sec:imp-notes}

A straightforward implementation following the conceptual description above leads to SMT timeouts on non-trivial examples.
The root cause is that the presentation writes behavioral predicates as IVL functions with explicit bodies ($\FUN \ldots = \mathit{body}$), and Boogie unfolds function bodies eagerly: any quantifier inside a body --- common in arithmetic identities or prelude vector axioms --- spawns trigger instantiations at every call site, causing combinatorial blowup in the SMT search.

We overcame this by adopting a pattern that decouples the \emph{definition} of behavioral predicates from their \emph{use}: declare each predicate as an \emph{uninterpreted} function and connect it to the per-function specification via axioms with explicit, carefully chosen triggers.
This pattern is not specific to higher-order functions; it is a general Boogie encoding discipline for modular specifications that keeps SMT instantiation under control.
The remaining implementation choices are natural refinements of applying this pattern consistently:

\begin{Itemize}
    \item \emph{Behavioral predicates are uninterpreted functions connected by axioms, not Boogie functions with explicit bodies.}
    The implementation declares $ensures_\tau$ as an uninterpreted Boogie function and connects it to the per-function spec via an axiom of the form $(\forall\, \ldots\; ::\;  \mathit{eval\_call} \iff \mathit{rhs})$ with an explicit trigger pattern.
    Axioms with explicit triggers fire only when the trigger pattern matches a ground term in the proof context, keeping instantiation under control.

    \item \emph{Per-variant trigger specialization.}
    The three variant kinds receive structurally different axioms.
    Shown for $ensures_\tau$ with the memory parameters elided for brevity, the axiom schemas are:

\begin{minipage}{\linewidth}
\begin{ivl}
\AXIOM \forall \sq{c}, \sq{p}, \sq{r} @ \{ensures_\tau(C_f(\sq{c}), \sq{p}, \sq{r})\} \\
\t1 ensures_\tau(C_f(\sq{c}), \sq{p}, \sq{r}) \iff ensures_{C_f(\sq{c})}(\sq{p}, \sq{r}) \\
\also
\AXIOM \forall \sq{p}, \sq{r} @ \{ensures_\tau(P_{f,x}(), \sq{p}, \sq{r})\} \\
\t1 ensures_\tau(P_{f,x}(), \sq{p}, \sq{r}) \iff ensures_{P_{f,x}}(\sq{p}, \sq{r}) \\
\also
\AXIOM \forall f, \sq{p}, \sq{r} @ \{ensures_\tau(f, \sq{p}, \sq{r})\} \\
\t1 f \IS F_{\sigma.x} \implies (ensures_\tau(f, \sq{p}, \sq{r}) \\
\t2 \iff ensures_{F_{\sigma.x}}(f.n, \sq{p}, \sq{r}))
\end{ivl}
\end{minipage}

    For closure variants, the trigger is the evaluator application \emph{containing} the constructor term $C_f(\sq{c})$, with the captures bound by the quantifier --- a bare constructor pattern would not cover the remaining bound variables; for parameter variants, the nullary ground constructor $P_{f,x}()$ plays the same role.
    For struct-field variants no such constructor term exists: at call sites a field-stored function value is an opaque datatype term loaded from memory.
    The trigger is therefore the evaluator application itself, with the variant pinned by the $f \IS F_{\sigma.x}$ guard and the discriminator passed via the selector $f.n$.
    In each case the axiom is instantiated only where the variant --- or, for fields, the evaluator application --- is syntactically present in the proof context.

    \item \emph{Constraining state-label conjuncts are dropped from the existential.}
    The conceptual behavioral predicate for an !aborts_if! at an intermediate state $\Mem^S$ includes \emph{all} ensures-with-$S$ conjuncts that reach $\Mem^S$ (including constraining ones such as $R[\Mem^S][a].0 > 0$).
    The implementation restricts the existential body to the \emph{defining} conjuncts --- those introduced by label-defining operations like $\ensuresof{f}{\bar{x}}$, $\resultof{f}{\bar{x}}$, or the publish/remove/update builtins.
    By the validity invariant $\bigvee \sq{A} \lor \bigwedge \sq{P}$, the constraining conjuncts hold at the witness chosen during verification of !foo!, so dropping them is sound and keeps the existential body small.

    \item \emph{State-label existential is flat, not recursive.}
    The recursive scheme above is logically equivalent to a single flat $(\exists\, \Mem^S, \Mem^T, \ldots\; ::\;  \ldots)$ wrapping the conjunction of defining fragments and the kind-specific clauses.
    The implementation uses the flat form, reducing the size of emitted Boogie and avoiding the additional SMT quantifier nesting that the recursive scheme would introduce.
\end{Itemize}

\begin{WideTable}
    \caption{Inference test categories that exercise behavioral predicates.}
    \label{tab:inference-tests}
    \centering
    \small
    \begin{tabularx}{\linewidth}{l X}
      \hline
      \textbf{Category} & \textbf{Representative cases} \\
      \hline
      Direct calls &
        Single and chained calls to a spec'd helper; inferred specs use
        \texttt{ensures\_of} and \texttt{aborts\_of} to abstract over the
        callee's effect rather than inlining its body. \\
      \hline
      Mutual recursion &
        Mutually recursive \texttt{is\_even}/\texttt{is\_odd}; inference
        breaks the cycle by phrasing each spec in terms of the other's
        behavioral predicate. \\
      \hline
      Higher-order parameters &
        A function that takes a closure and applies it; the inferred spec
        abstracts the closure via \texttt{ensures\_of}/\texttt{aborts\_of}
        on the parameter, with no commitment to a particular implementation. \\
      \hline
      Stored-closure dispatch (enum) &
        Calculator with an enum variant carrying a closure field;
        dispatch through the variant is summarized by behavioral
        predicates on that field. \\
      \hline
      Stored-closure dispatch (struct field) &
        Vault with a \texttt{Strategy} closure stored in a struct field on
        Aptos framework types; inference emits \texttt{modifies\_of} for
        the field together with the field-form behavioral predicates. \\
      \hline
      Pre/post state labels &
        \texttt{move\_to}, \texttt{move\_from}, mutable resource indexing;
        inferred specs distinguish \texttt{@pre} from post-state on the
        memory the predicate observes. \\
      \hline
      Intermediate state labels &
        Sequenced state-modifying calls (e.g.\ remove-then-publish);
        inferred specs introduce existentially quantified intermediate
        labels and pin them to defining behavioral predicates. \\
      \hline
    \end{tabularx}
\end{WideTable}

\Section{Specification Inference}
\label{sec:inference}

\SubSection{How it Works}
\label{sec:validation-overview}

\MVP's \emph{specification inference} \cite{GrieskampEtAl26Inf} based on weakest-precondition (\WP) analysis~\cite{dijkstra1975wp,BL05} is made feasible and modular via behavioral predicates.
\WP inference takes arbitrary Move code and, for each function, computes conditions that are both necessary and sufficient for the implementation's behavior.
Spec inference uses behavioral predicates to express the effect of calling out to another function, whether through a function value or a direct call.
It reverses the translation seen for $invoke_\tau$ in Sec.~\ref{sec:enc-invoke}, introducing the behavioral predicates into the \WP to express the call. Without this modular abstraction, inferred specs would be impractically large.

A full description of \MVP's \WP analysis is out of scope of this paper. What is important is that, besides the notorious loop invariant problem for \WP, specification inference for Move is precise and complete.
This serves two purposes for this work, demonstrated in the following subsections: reducing the specification burden for compliance checking of function values (Sec.~\ref{sec:validation-amm}), and providing a validation pipeline that confirms the encoding is internally consistent (Sec.~\ref{sec:validation-tests}).

\SubSection{Compliance Checking}
\label{sec:validation-amm}

Recall from Sec.~\ref{sec:amm} that the !Pool! struct carries a !pricing! closure constrained by a no-abort invariant: the pricing function must not abort for any inputs in any state.
Two concrete pricing implementations were shown: !product!, which is compliant (it never aborts), and !with_fee!, which is not (it aborts when the owner's !Fee! resource is absent).
Spec inference makes this compliance distinction machine-readable without any manual annotation.

Given the implementation of !with_fee!, \WP inference easily derives the aborts conditions (we leave out the ensures, which it also determines):

\begin{MoveBox}
  spec with_fee {
    aborts_if [inferred] !exists<Fee>(owner);
    aborts_if [inferred] Fee[owner].bps > 10000;
    ...
  }
\end{MoveBox}

Similarly, for the lambda example in Sec.~\ref{sec:find}, we were required to attach a spec to a trivial lambda expression of the form !|x| x == 0!.
Obviously, inference can help to avoid this.
We are currently investigating relaxing the requirement of spec blocks for compliance-checked functions in those cases where they can be reliably inferred (no recursion and no loops); however, this is orthogonal to the work in this paper.

\SubSection{Validation Pipeline and Test Coverage}
\label{sec:validation-tests}

Inferred conditions are labeled \texttt{[inferred]} and feed directly back into \MVP for verification, creating a pipeline: code $\leadsto$ code+specs $\leadsto$ VC, with verification expected to succeed.
This pipeline acts as a \emph{differential-testing oracle} for the extension described in this paper: inference and verification are independent implementations which meet only at the semantics of behavioral predicates and state labels, so a failure indicates a defect in one of them --- conditions inferred unsoundly or not tightly, or axioms too weak to discharge them.
Tab.~\ref{tab:inference-tests} maps the categories of the inference test suite~\cite{InferenceTestsuite} --- currently 27 cases with over 550 inferred conditions --- to the HOF patterns introduced in this paper.
All categories pass end-to-end, evidence that the encoding is internally consistent and that the axioms are strong enough to discharge \emph{tight} specifications: usable, not merely sound.
Mutual recursion deserves a note: inference breaks the specification cycle by phrasing each function's spec in terms of the other's behavioral predicate, which verification resolves against the axioms of Sec.~\ref{sec:enc-bp}.

\SubSection{Empirical Status}
\label{sec:empirical}

The extension described in this paper is implemented in the production \MVP shipped with the Aptos toolchain; the paper describes Move language version 2.4 and \MVP as of the aptos-core commit pinned in~\cite{PaperExamples, InferenceTestsuite}.
The closure portion of the prover's test suite comprises 19 test modules with about 240 verified functions and 66 expected-error diagnoses, including the examples of this paper; together with the inference suite, these run in continuous integration on every change to the prover.
The tests are required to be stable across changes to the prover and underlying tools like Z3, and complete in under 5 minutes on standard GitHub runners, with a default timeout of 40 seconds per verification condition.
A benchmark over a larger corpus of higher-order Move contracts does not yet exist: higher-order Move shipped in late 2025, and a corpus is only now forming, partly in closed-source projects.


\Section{Related Work and Conclusion}
\label{sec:conclusion}

\SubSection{Related work}
\label{sec:related}

Higher-order programs with side effects have been a long-standing concern in program verification.
Our approach to behavioral predicates descends from the relational tradition of VDM~\cite{jones1990vdm}, Z~\cite{spivey1992z, grieskamp2000executablez}, and the B method~\cite{abrial1996bbook}, in which operations are specified by relations between pre- and post-states; state labels generalize this to intermediate snapshots and make the relational flavor composable.

Among contemporary verifiers for imperative code with higher-order functions, \verus~\cite{verus} offers function-value specifications via \emph{spec closures} and phantom ghost state; the approach is state-passing, with state threaded as an extra argument to specifications rather than quantified over directly.
Our treatment is more directly relational: state labels denote memory snapshots and are bound by quantifiers in the surface syntax, avoiding the encoding cost of an explicit state argument and matching the style in which Move developers already write pre-/postcondition specifications.
Also for Rust, Wolff et al.~\cite{RustClosures21} verify closures in the Prusti verifier via \emph{call descriptions}, which relate a closure call's arguments and results to its captured state within separation logic; our behavioral predicates play a similar role, but are relational over labeled memory snapshots, with Move's alias-free references and resource-separated memory removing the need for separation-logic machinery.

\fstar~\cite{fstar2016} likewise verifies higher-order, effectful code, internalizing the WP calculus in its type system via Dijkstra monads~\cite{dm4all}.
Both systems support pre-/postcondition specifications, quantifiers, behavioral abstraction over function values, and structured proof guidance --- in Move via \texttt{proof} blocks, \texttt{lemma} declarations, and \texttt{calc}, \texttt{split}, \texttt{apply} statements.
The treatment of memory differs: \fstar~optionally supports separation logic (via the Steel and Pulse frameworks) to reason about heap separation within the logic, whereas Move uses traditional frame conditions via modifies declarations, and its resource and reference model separates memory syntactically at compile time --- separation need not be reasoned about by the SMT solver.

Dafny~\cite{DAFNY} is a close cousin of \MVP: it has a similar specification model of pre-/postconditions, higher-order functions, and behavioral predicates.
However, Dafny does not allow function values to modify state.
This would be a severe restriction for Move, since it is a common usage pattern.
Moreover, Move's resource-separated global memory and alias-free mutable reference model is not available in Dafny, as is also the case in \fstar.
While both languages support state labels, Move state labels as described here can be used in abstract specifications via universal and existential quantification.

Lean~4~\cite{lean4} is a dependently-typed interactive theorem prover~\cite{Isabelle, COQ} supporting higher-order functions, with tactic-driven, kernel-checked verification.
Unlike \MVP, Lean does not work directly on a programming language but requires the user to encode it in its meta logic, typically modeling stateful systems via monads.
\MVP, in contrast, integrates directly with Move and aims at automated SMT-based verification, with optional proof scripts.

In the smart-contract setting, dynamic dispatch is a central challenge for \solidity verifiers.
\certora~\cite{certora} handles dispatch on Solidity by enumerating the possible callees of a virtual call (derived from the contracts in scope) and generating a case-split proof obligation per callee; summaries can be supplied by the user to abstract external calls.
Other SMT-based Solidity tools~\cite{smtalt18,solcverify,DBLP:conf/esop/HajduJ20} handle dynamic dispatch through similar case-splitting, with varying degrees of abstraction.
\MVP is more general: a call site's dispatch is explicit in the closure's datatype, and our encoding lets the prover automatically fall back from case-split (when the concrete function is known) to behavioral-predicate reasoning (when only the function type is known), without the user having to structure the specification differently.

\SubSection{Future Work}

The current encoding enumerates all closure variants reaching a given call site; scaling to programs where the variant set is large or open (e.g.\ dynamically loaded modules) remains future work.
State labels are currently individually bound; extending quantification to \emph{sequences} of states ($\forall S_0..S_n$) would allow direct contracts for fold-like iteration, where today the higher-order function is verified once against a summary spec (Sec.~\ref{sec:find}) --- the flat existential encoding of Sec.~\ref{sec:enc-frame} already handles chains of fixed length and would extend naturally.
Proof automation for non-linear arithmetic arising in pricing formulas and similar domains currently relies on user-supplied proof hints; integrating portfolio arithmetic decision procedures is a natural direction.
The spec inference pipeline produces behavioral-predicate-based specifications from arbitrary code; extending it to synthesize loop invariants involving function values automatically would reduce the remaining manual annotation burden.

\SubSection{Conclusion}

We extended \MVP to Move 2's higher-order functions and dynamic dispatch.
The extension rests on two language additions --- behavioral predicates and state labels --- and on a Boogie encoding that discriminates closure invocations on the function-value variant reaching the call site.
The design preserves \MVP's production-oriented cost model: routine, predictable, counterexample-driven verification, also in the presence of dynamic dispatch.
We further extended \MVP's specification inference to closures and dynamic dispatch, turning compliance of a function value with a behavioral invariant into an automated verification problem.


\newpage
\bibliographystyle{IEEEtran}
\bibliography{biblio}

\end{document}